\documentclass{emulateapj}
\slugcomment{draft v1}

\shorttitle{X-ray counterparts to gravitational waves}
\shortauthors{Kisaka, Ioka \& Nakamura}

\begin{document}


\title{Isotropic detectable X-ray counterparts to gravitational waves from neutron star binary mergers}


\author{Shota Kisaka\altaffilmark{1}}
\email{kisaka@post.kek.jp}
\author{Kunihito Ioka\altaffilmark{1,2}}
\email{kunihito.ioka@kek.jp}
\author{Takashi Nakamura\altaffilmark{3}}
\email{takashi@tap.scphys.kyoto-u.ac.jp}


\altaffiltext{1}{Theory Center, Institute of Particle and Nuclear Studies, KEK, Tsukuba 305-0801, Japan}
\altaffiltext{2}{Department of Particle and Nuclear Physics, SOKENDAI (The Graduate University for Advanced Studies), Tsukuba 305-0801, Japan}
\altaffiltext{3}{Department of Physics, Kyoto University, Kyoto 606-8502, Japan}


\begin{abstract}
Neutron star binary mergers are strong sources of gravitational waves (GWs).
Promising electromagnetic counterparts are short gamma-ray bursts (GRBs) 
but the emission is highly collimated.
We propose that the scattering of the long-lasting plateau emission in short GRBs by the merger ejecta
produces nearly isotropic emission for $\sim 10^4$ s
with flux $10^{-13}-10^{-10}$ erg cm$^{-2}$ s$^{-1}$ at 100 Mpc in X-ray.
This is detectable by {\it Swift XRT} and wide field X-ray detectors such as
{\it ISS-Lobster, Einstein Probe, eROSITA} and {\it WF-MAXI},
which are desired by the infrared and optical follow-ups to localize 
and measure the distance to the host galaxy.
The scattered X-rays obtain linear polarization, which correlates with
the jet direction, X-ray luminosity and GW polarizations.
The activity of plateau emission is also
a natural energy source of a macronova (or kilonova) detected in short GRB 130603B
without the $r$-process radioactivity.
\end{abstract}





\section{INTRODUCTION}

Electromagnetic counterparts to gravitational wave (GW) sources are important to maximize scientific returns from the detection of GWs \citep[e.g., ][]{MB12}.
One of the most promising candidates for the direct detection of the GW is a merger of a neutron star (NS) binary\footnote{We use a term ``a binary NS" for a NS-NS binary and ``a NS binary" for a NS-NS or black hole-NS (BH-NS) binary.}.
There are several models proposed for the electromagnetic counterparts of the binary mergers \citep[e.g., ][]{R15}.

Short gamma-ray bursts (GRBs) are considered as an electromagnetic counterpart of the NS binary merger \citep[e.g., ][]{B14}. 
The light curve of short GRBs shows several components \citep[e.g., ][]{Fox+05}.
Initial gamma-ray spikes are prompt emissions with the luminosity $\sim10^{50}-10^{51}$ erg s$^{-1}$ \footnote{To be precise, \citet{yonetoku2014} suggest that the cumulative luminosity function is inversely proportional to the luminosity in the range of $10^{50}-10^{53}$ erg s$^{-1}$.} and duration $\sim10^{-1}-1$ s. 
These are followed by the extended emission \footnote{We define the extended emission as the emission for $\sim10^2$ s, which also includes the plateau component analyzed by \citet{Row+13} and \citet{Lu+15}. Note that the fraction of the short GRB with the extended emission could be significantly larger in softer energy bands \citep{Nak+14}.} 
\citep[e.g., ][]{Bar+05,Kag+15}, which has the luminosity $\sim10^{48}-10^{49}$ erg s$^{-1}$ and duration $\sim10^2$ s.
These components show a sharp drop in the light curve, which cannot be produced by the afterglow and hence requires the activities of the central engine \citep{IKZ05}.
The origin of the emissions is most likely a collimated relativistic jet. 

Some short GRBs show a long-lasting plateau component\footnote{Note that the plateau emission would be sometimes hidden by the afterglow emission or below the detection limit. These events may correspond to the ``no breaks'' in \citet{Row+13} and ``no plateau samples'' in \citet{Lu+15}.} with 
the luminosity $\sim10^{46}-10^{47}$ erg s$^{-1}$ and 
duration $\sim10^3-10^4$ s in their light curves \citep{Row+13, GOWR13}.
Thus the fluence of each of three components is of roughly the same order of magnitude \citep{Row+13}.
The plateau emission is considered to be produced by an activity of the central engine such as a relativistic jet from a BH with a typical NS magnetic field $\sim10^{12}$ G \citep{KI15} or a pulsar wind from a highly magnetized ($\sim10^{15}-10^{16}$ G) and rapidly rotating ($\sim1$ ms) NS \citep{Fan+13, Row+13, GOWR13}.

Nearly isotropic emissions from a merger event have been anticipated (e.g., \citealt{LP98}, \citealt{K05}, \citealt{KBB13}, \citealt{TH13}, \citealt{TNI14}, \citealt{Nak+14}, \citealt{KIT15}, hereafter KIT15) because of the high probability of the simultaneous detection with GWs.
Recently, numerical simulations suggest that significant mass is isotropically ejected by a NS binary merger \citep[e.g., ][]{Hot+13, Kyutoku:2013wxa, Kyu+15}\footnote{Although the dynamical ejecta expand with significant anisotropy in a BH-NS merger, some ejecta blow into the direction of the binary orbital axis \citep[e.g., ][]{Kyutoku:2013wxa,Kyu+15}. The disk wind also ejects the mass to the rotation axis of the disk \citep[e.g., ][]{Fer+15}. }. 
The collimated outflow from the central engine due to the Blandford-Znajek process \citep{Blandford1977} or the pulsar wind interacts with the isotropic ejecta and emits the isotropic radiation.

In this Letter, we consider nearly isotropic emissions caused by the long-lasting activity, which produces a plateau emission, as electromagnetic counterparts of the NS binary merger. 
In particular, we focus on a scattering of plateau X-ray photons \citep[e.g., ][]{N98, EL99}. We also consider a macronova\footnote{We use the term ``macronova'' as a transient with a NS binary merger, especially thermal radiation from the merger ejecta.} (or kilonova) powered by the plateau activity (KIT15),
without resort to the $r$-process radioactivity \citep{Tan+13,BFC13}.
These detections would significantly reduce the localization error of GW detectors \citep[$\sim10-100$ deg$^2$; e.g., ][]{Ber+15}.

The Letter is organized as follows: in Section 2 we estimate the luminosity of the scattered plateau emission, and compare it with the sensitivity of X-ray observations. 
In Section 3, we present the model of a macronova powered by the plateau activity, which explains the observations of GRB 130603B. 
Finally, we present discussions in Section 4. 

\section{Scattered X-ray emission}

\begin{figure}
 \begin{center}
  \includegraphics[width=70mm]{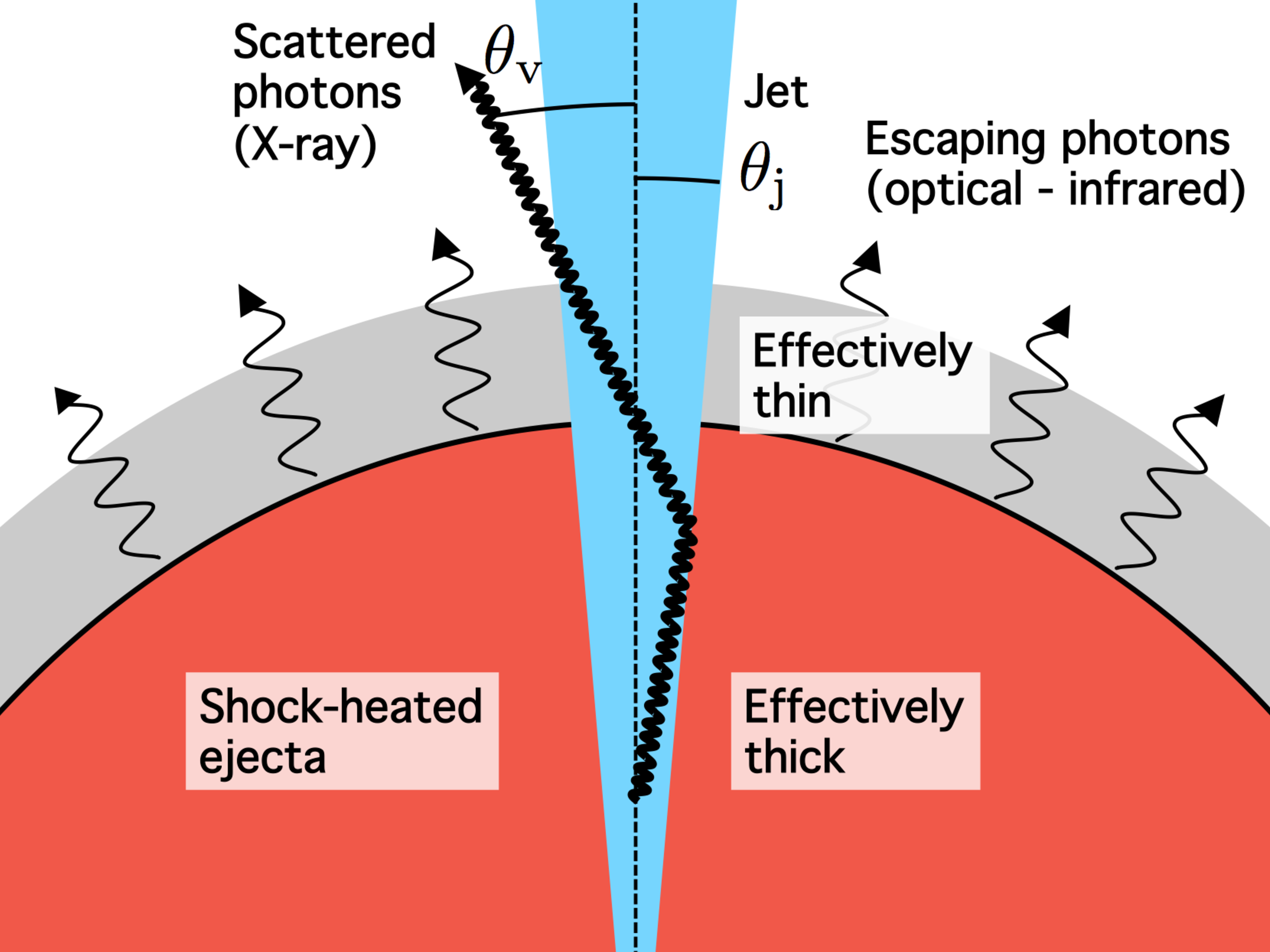}
   \caption{Schematic picture for the scattering of plateau emission and the engine-powered macronova. X-ray photons emitted from the inside of the jet (light blue region) are scattered by the optically thick ejecta (thick arrow). The grey region is effectively thin and the red region is effectively thick.}
  \label{figure:mn}
 \end{center}
\end{figure}

\begin{figure}
 \begin{center}
  \includegraphics[width=60mm, angle=270]{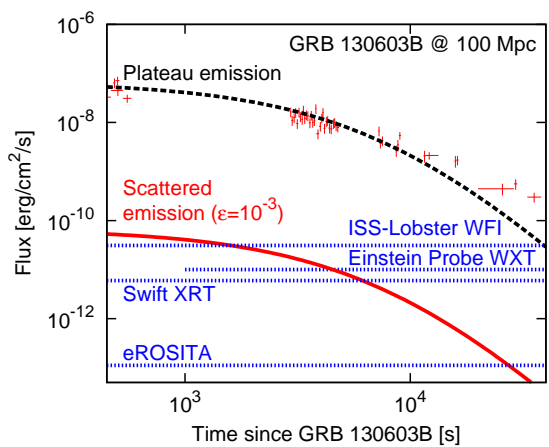}
   \caption{Light curves of the plateau (the black dashed curve) and its scattered emissions ($\epsilon=10^{-3}$; the red solid curve). 
Red crosses are the plateau emission of GRB 130603B with the distance changed from the original redshift $z=0.356$ to 100 Mpc.
Observational data are obtained from UK {\it Swift} Science Data Centre. Blue dotted lines show the sensitivity limits for the soft X-ray detectors of {\it ISS-Lobster}/WTI (integration time 450 s), {\it Einstein Probe}/WXT (integration time 1000 s), {\it Swift}/XRT (integration time 100 s) and {\it eROSITA} (integration time corresponding to a single survey pass). The scattered emission is detectable for these X-ray detectors.}
  \label{figure:x}
 \end{center}
\end{figure}

Figure \ref{figure:mn} shows a schematic picture for the scattering of the emission from the jet (the thick arrow). 
A significant fraction of photons which are emitted with angle $\gtrsim\theta_{\rm j}$ relative to the jet axis could be scattered at a large angle by the surrounding ejecta if the optical depth for the Thomson scattering is larger than unity, $\tau\sim n\sigma_{\rm T} r\gg 1$, where $n$ is the electron number density, and $\sigma_{\rm T}$ is the Thomson cross section.
Using the assumption of homologous expansion for the ejecta \citep{Hot+13}, the radius of the ejecta $r$ is described by the velocity $v$ and time since the merger $t$ as $r\sim vt$. 
The number density\footnote{Typically the nuclei are weakly ionized.} is described by $n\sim M_{\rm ej}/(\bar{A}m_{\rm p}v^3t^3)$, where $\bar{A}$ is the average mass number of the nuclei in the ejecta and $m_{\rm p}$ is the proton mass.
If the ejecta mainly consist of the $r$-process elements, we have $\bar{A}\sim100$ \citep[e.g., ][]{LS74}.
Then a typical value of the optical depth is 
\begin{eqnarray}
\tau\sim10^2\left(\frac{t}{10^4\,{\rm s}}\right)^{-2}\left(\frac{\bar{A}}{10^2}\right)^{-1}\left(\frac{M_{\rm ej}}{10^{-2}\,M_{\odot}}\right)\left(\frac{v}{0.1\,c}\right)^{-2},
\end{eqnarray}
where $c$ is the speed of the light. 
Therefore, the surrounding ejecta are optically thick to the Thomson scattering during the plateau activity timescale ($\sim10^4$ s).

Another condition to scatter a significant fraction of the plateau emission is that the radius of the plateau emission region is smaller than that of the expanding ejecta (Figure \ref{figure:mn}). 
Since the typical velocity of the ejecta is $v\sim0.1c$, the radius of the ejecta is described by
\begin{eqnarray}
r\sim 3\times 10^{13} \left(\frac{v}{0.1~c}\right)\left(\frac{t}{10^4\,{\rm s}}\right)~~{\rm cm}.
\end{eqnarray}
On the other hand, the radius of the plateau emission region is estimated as 
\begin{eqnarray}
r_{\rm plateau}\sim \Gamma^2c\Delta t 
\sim 3\times 10^{12} \left(\frac{\Gamma}{10}\right)^2\left(\frac{\Delta t}{1\,{\rm s}}\right)~~{\rm cm},
\end{eqnarray}
where $\Gamma$ is the bulk Lorentz factor of the emitter and $\Delta t$ is the flux variability timescale. 
Both $\Gamma$ and $\Delta t$ have some range for each event, so that the emission continues over $r_{\rm plateau}\sim10^{11}-10^{14}$ cm in an approximately logarithmic way.
Since the Lorentz factor is low $\Gamma \sim 10$ inside the jet due to the cocoon confinement
\citep{Nag+14} and thus the relativistic beaming angle is larger than the jet opening angle
$1/\Gamma \gtrsim \theta_j$, 
most emission from the jet can reach the boundary between the jet and the ejecta as shown in Figure \ref{figure:mn}.
In addition, the range of $r_{\rm plateau}$ covers the sweet spot for the scattering, $r_{\rm plateau}\sim r/\Gamma \sim 3\times10^{12}\,{\rm cm}$.
Therefore, the observers with large viewing angle ($\gtrsim\theta_{\rm j}$) with respect to the jet axis would detect the scattered X-ray photons of the plateau emission.

We parameterize the scattered luminosity of the plateau emission $L_{\rm rf}$ using a parameter $\epsilon$ as
\begin{eqnarray}
L_{\rm rf}\sim\epsilon L_{\rm iso,pl},
\end{eqnarray}
where $L_{\rm iso,pl}$ is the observed isotropic luminosity of the plateau emission.
We first consider the isotropically scattered component whose energy is comparable to that before the scattering.
Then, the luminosity of the scattered component is 
$L_{\rm rf}\sim[(\theta_{\rm j}^2/2)]L_{\rm pl}\sim5\times10^{-3}(\theta_{\rm j}/0.1)^2L_{\rm pl}$. 
Taking account of the widespread region of the emission, we use $\epsilon=10^{-3}$ as a fiducial case. 
Some geometrical models suggest $\epsilon\sim3\times10^{-5}-3\times10^{-4}$ in the case of $r_{\rm plateau}/r=0.1$ \citep[e.g., Equation (3) in ][]{EL99}, which is a bit smaller than the fiducial value.
Note that the light crossing time of the emission region, $\theta_j r/c \sim 100$ s, is smaller than the plateau duration, so that the luminosity is not reduced by the time stretch due to the light crossing.

In Figure \ref{figure:x}, we plot the light curves of the plateau emission (the black dashed curve) and the scattered component with $\epsilon=10^{-3}$ (the red solid curve).
A black dashed curve is the model light curve \citep[Equation (12) in ][]{KI15} with the luminosity $L_{\rm iso,pl}=8\times10^{46}$ erg s$^{-1}$ and duration $t_{\rm inj}=8.5\times10^3$ s.
We also plot the flux (0.3-10 keV) of the plateau emission\footnote{http://www.swift.ac.uk/index.php} assuming that GRB 130603B like event occurs at the distance 100 Mpc (red crosses). 
The photon index of the plateau emission is about $\sim-2$, so that the flux of the plateau emission does not strongly depend on the energy range.
To see the detectability, we also plot the sensitivities of some soft X-ray detectors (blue dotted lines).

As shown in Figure \ref{figure:x}, the flux of the scattered component with $\epsilon\sim10^{-3}$ at 100 Mpc is comparable to the flux sensitivity limit of {\it ISS-Lobster}/WTI with integration time 450 s \citep{Camp+13} and {\it Einstein Probe}/WXT with integration time 1000 s \citep{Yuan+15}. 
{\it ISS-Lobster}/WTI and {\it Einstein Probe}/WXT can cover a wide field of 900 deg$^2$ and 3600 deg$^2$ per pointing, respectively.
The pointing observations by these detectors can detect the X-ray counterpart to the NS binary merger within $\sim10^3-10^4$ s after the GW alert. 

We also show the sensitivity limit of {\it eROSITA} in Figure \ref{figure:x}. 
The survey flux sensitivity is about $\sim10^{-13}$ erg cm$^{-2}$ s$^{-1}$ \citep{Mer+12}, so that the scattered component with even $\epsilon\sim10^{-4}$ can be detected at 100 Mpc (Figure \ref{figure:x}).
The {\it eROSITA} field of view is 0.833 deg$^2$. 
The scan area of {\it eROSITA} during the plateau emission $\sim10^4$ s is about $\sim0.5\%$ of the whole sky \citep{Mer+12}. 
Since the detection horizon of the GW detectors such as Advanced LIGO, Advanced VIRGO and KAGRA is $\sim200$ Mpc, the expected rate of the simultaneous detection by {\it eROSITA} survey and GW detectors is $\sim4\times10^{-2}~(R_{\rm merger}/10^3~{\rm Gpc}^{-3}{\rm yr}^{-1})$ yr$^{-1}$ where $R_{\rm merger}$ is the rate of the NS-NS merger \citep[e.g., ][]{Aba+10}. 
If the merger rate is $\sim6$ times larger than the canonical value, $R_{\rm merger}\sim6\times10^3~{\rm Gpc}^{-3}{\rm yr}^{-1}$, we expect $\sim1$ simultaneous detection during the four years all-sky survey.

{\it Swift}/XRT can detect the scattered component of the plateau emission even if the scattering parameter is $\epsilon\sim10^{-4}$.
We also plot the flux sensitivity line of the pointing observation by {\it Swift}/XRT with integration time $10^2$ s \citep{Kan+12}.  
Although field of view of {\it Swift}/XRT, 0.16 deg$^2$, is much smaller than the GW error box, the galaxy catalog within $\sim 100$ Mpc could make it possible to detect the scattered component (see Section 4).

\section{Macronovae}

\begin{figure}
 \begin{center}
  \includegraphics[width=60mm,angle=270]{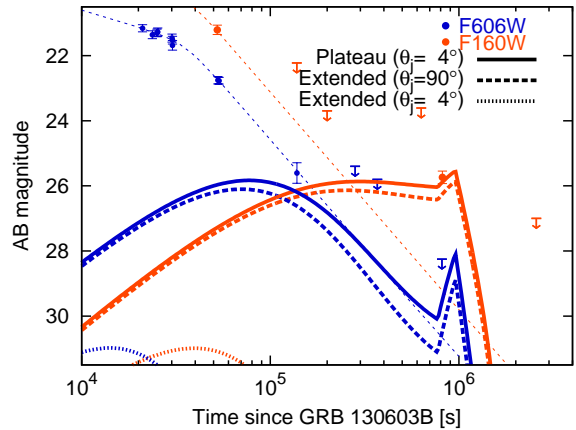}
   \caption{Theoretical light curves of engine-powered macronovae at the near-infrared (F160W, red) and optical bands (F606W, blue). We consider three models of the energy sources, plateau emission with $\theta_{\rm j}=4^\circ$ (solid line), extended emissions with $\theta=90^\circ$ (dashed line) and $\theta_{\rm j}=4^\circ$ (dotted line). The observational results of GRB 130603B \citep[$z = 0.356$; ][]{Cuc+13, Tan+13, BFC13, de+14} are also plotted. The thin dotted lines are light curves calculated from a GRB afterglow model \citep{Tan+13}. Models of the collimated plateau emission ($\theta_{\rm j}=4^\circ$) and the isotropic extended emission ($\theta_{\rm j}=90^\circ$) reproduce the observational data well.}
  \label{figure:opt}
 \end{center}
\end{figure}

In Figure \ref{figure:mn}, we show a schematic picture for the engine-powered macronova. 
We compare two heating sources, activities of the plateau and extended emissions.
The light curve modeling of an engine-powered macronova is the same as the engine model in KIT15. 
We use Equations (A.13) and (A.15) in Appendix A of KIT15 to describe the macronova light curves.
For the model parameters, we choose the ejecta mass $M_{\rm ej}=0.1M_{\odot}$, the maximum and minimum velocities of the ejecta $v_{\min}=0.15c$ and $v_{\max}=0.4c$, the index of the mass density profile $\beta=3.5$, the opacity of the ejecta $\kappa=10$ cm$^2$ g$^{-1}$ and the index of the temperature distribution $\xi=1.6$, which are the same as the fiducial model in KIT15.
The remaining model parameters are the injection timescale $t_{\rm inj}$ and the injected energy $E_{\rm int0}=[(\theta_{\rm j}^2/2)/\eta]L_{\rm iso}t_{\rm inj}$, which are determined by the radiative efficiency $\eta=0.1$ \citep[e.g., ][]{Zha+07}, the jet half-opening angle $\theta_{\rm j}$, the observed isotropic luminosity $L_{\rm iso}$ and the duration of the central engine activity. 
Note that 
the engine power is alternative to the radioactive heating \citep[e.g., ][]{LP98}.

First, we consider the plateau activity as a heating source of the ejecta. From the observations of GRB 130603B, the observed isotropic luminosity and the duration of the plateau emission are $L_{\rm iso,pl}\sim8\times10^{46}$ erg s$^{-1}$ and $t_{\rm inj}\sim8.5\times10^3$ s, respectively (see Figure \ref{figure:x}). 
Using the measured half-opening angle $\theta_{\rm j}=4^{\circ}$ \citep{Fong+14}, the injected energy is $E_{\rm int0}=1.6\times10^{49}$ erg  \footnote{To be precise, the adiabatic cooling in the early phase
reduces the energy by $\sim 0.1$,
although this may be absorbed by the uncertainties of the other parameters.
}.

Second, we consider the activity of the extended emission as a heating source of the ejecta. Since the extended emission was not detected in GRB 130603B, we assume the typical extended emission, $L_{\rm iso,ee}=10^{48}$ erg s$^{-1}$ and $t_{\rm inj}=10^2$ s.
Then, the injected energy is $E_{\rm int0}=2.5\times10^{49}$ erg for the half-opening angle of the extended emission $\theta_{\rm j}=4^{\circ}$. 
For comparison, we also consider the isotropic extended emission ($\theta_{\rm j}=90^{\circ}$) as an extreme case. 
Then, the injected energy is $E_{\rm int0}=10^{51}$ erg.

In Figure \ref{figure:opt}, we compare the model light curves with the observations of GRB 130603B \citep[$z=0.356$; ][]{Cuc+13, Tan+13, BFC13, de+14}. 
Macronovae powered by the activities of the collimated plateau emission ($\theta_{\rm j}=4^{\circ}$; thick solid lines) and the isotropic extended emission ($\theta_{\rm j}=90^{\circ}$; thick dashed lines) are almost consistent with the observations.
However, the model of the collimated extended emission ($\theta=4^{\circ}$; thick dotted lines) is too dim to reproduce the observations. 
Since both the observed luminosity and the temperature of the macronova are described by $L_{\rm bol}\propto E_{\rm int0}t_{\rm inj}$ and $T_{\rm obs}\propto(E_{\rm int0}t_{\rm inj})^{1/4}$, respectively, the longer activity is more important for the brightness of the engine-powered macronova. 
This is due to the effect of the adiabatic cooling.
The plateau emission continues for a longer time and thereby more naturally
reproduces the observed infrared excess than the extended emission (KIT15).

\section{DISCUSSIONS}

We investigate the isotropic electromagnetic emission from a NS binary merger, which caused by the long-lasting plateau activity associated with short GRBs.
In particular, we focus on the scattering of the X-ray photons by the ejecta and the engine-powered macronova.

We suggest that the scattered X-ray component of the plateau emission could be detected by the future soft X-ray experiments. 
The luminosity, $\sim10^{44}$ erg s$^{-1}$ ($\epsilon=10^{-3}$), is as bright as the X-ray break luminosity of the AGN luminosity function at redshift $z\lesssim1$. 
This is also brighter than the previous models of isotropic X-ray counterparts at a follow-up time $\sim 10^{3}$--$10^{4}$ s, such as the ultrarelativistic shock \citep{KIS14} and the merger ejecta remnant \citep{TKI14} (but see also \citealt{Zha+13} and \citealt{Nak+14}).
The follow-up observations in X-ray band are essential to localize the GW sources because X-ray sources detected within $\sim 0.1^\circ$ error box can make it possible to perform further optical and infrared follow-ups to identify the host galaxy.
The detection of the scattered component gives not only the localization of the GW sources with $\lesssim0.1^\circ$, but also the hints for the long-lasting central engine. 

The flux of the scattered plateau emission with $\epsilon\sim10^{-3}$ at 100 Mpc is comparable to the sensitivity of the {\it ISS-Lobster}/WFI with integration time 450 s.
Although {\it ISS-Lobster} is planning to survey for 50 s in each field of view until the GW position is uploaded \citep{Camp+13},
we suggest more integration time than $\gtrsim450$ s and hopefully $\sim$2000 s 
to detect the scattered component down to $\epsilon\sim10^{-4}$ at 100 Mpc.
After the GW telescopes localize the source, {\it ISS-Lobster} will start the pointing observation \citep{Camp+13}.
{\it ISS-Lobster}/WFI has the wide field of view (900 deg$^2$), which is larger than the localization of the GW error box ($\sim100$ deg$^2$). 
Since 
the GW error box for the localization is 
worse with fewer GW detectors
and also highly elongated \citep[e.g., ][]{Ber+15},
several pointings may be necessary to encompass the GW error box.

As mentioned in Section 2, the detection rate of the scattering X-ray emission by {\it eROSITA} within the detection horizon of the GW detectors ($\sim$200 Mpc) is $\sim4\times10^{-2}(R_{\rm merger}/10^3{\rm Gpc}^{-3}{\rm yr}^{-1})$ yr$^{-1}$. 
However, since {\it eROSITA} could detect the scattered component with $\epsilon\sim10^{-3}$ up to the distance $\sim$1 Gpc, the expected detection rate 
only by {\it eROSITA} survey observation 
is $\sim6~(R_{\rm merger}/10^3~{\rm Gpc}^{-3}{\rm yr}^{-1})~{\rm yr}^{-1}$. 
Therefore, we suggest that the combination with the detection by {\it eROSITA} and the follow-up observations in optical and infrared bands within $\lesssim10$ days could identify the merger events without GW detections. 

We suggest that the follow-up observations by {\it Swift}/XRT 
towards the galaxies in the GW error box ($\sim 100$ deg$^{-2}$) could detect the scattering components when the GW detection horizon is $\sim 100 $ Mpc
\citep{Kan+12}. 
The number of the galaxies within $100$ Mpc in the GW error box is $\sim 10^2$ since the number density of the galaxy is $\sim 10^{-2}$ Mpc$^{-3}$. 
Then, {\it Swift}/XRT could observe these galaxies within the plateau timescale $\sim 10^4$ s with integration time $\sim10^2$ s for each galaxy.

The scattering model is consistent with {\it XMM-Newton} Slew Survey \citep{Kan+13}, which detected 6 soft X-ray transients.
They are spatially coincident with previously cataloged galaxies within 350 Mpc, lack evidence for active galactic nuclei, and display luminosities $\sim10^{43}$ erg s$^{-1}$ (corresponding to $\epsilon\sim10^{-3}$).
If the duration is $T_{\rm dur}\sim10^4$ s, the event rate is $2\times10^4(T_{\rm dur}/10^4~{\rm s})^{-1}$ yr$^{-1}$ ($<300$ Mpc), larger than that of the binary NS mergers $\sim 30(R_{\rm merger}/10^3{\rm Gpc}^{-3}{\rm yr}^{-1})$ yr$^{-1}$ ($<300$ Mpc)\footnote{
Note that this large difference of event rates may not indicate that most of transients detected by {\it eROSITA} survey will not be NS binary mergers. 
Tidal disruption events are considered as candidates of X-ray transients \citep{Kan+13}. 
Under the assumptions of an initial luminosity $\sim10^{45}$ erg s$^{-1}$ during few days or weeks followed by a characteristic dimming $\propto t^{-5/3}$, the timescale of such an event with luminosity $\sim10^{43}$ erg s$^{-1}$ is $T_{\rm dur} \sim 10^6 - 10^7$ s \citep{Kan+13}. 
Then, if {\it XMM-Newton} Slew Survey transients are tidal disruption events, the event rate is $20 (T_{\rm dur}/10^7{\rm s})^{-1}$ yr$^{-1}$ ($<$300 Mpc), which is comparable to that of the binary NS merger.}.

We briefly discuss about the anisotropy of the scattered component. 
In Section 2.1, we consider a model that the collimated plateau emission is scattered into $4\pi$ direction. 
In the geometrical model of \citet{EL99}, the intensity of the scattered component depends on the emission direction, being weaker for larger viewing angle (see their Figure 2).
Moreover, if the ejecta have a large velocity ($v\sim c$) by such as the central engine activity, the scattered photons become more anisotropic.
The anisotropy affects the event rate of the simultaneous detections of the GW and X-ray emission from NS binary mergers.
The large velocity of the ejecta produces a macronova with short duration (KIT15) and the bright radio flare at late time \citep[$\gtrsim1$ yr; e.g., ][]{NP11, PNR13, TKI14}. 
Therefore, the simultaneous observations in GW, X-ray, optical/infrared and radio bands give the detail information about the structure of the ejecta and the activity of the central engine.

The scattered X-ray is linearly polarized, so that it brings new information.
First the polarization degree $\Pi=(1-\cos^2\theta)/(1+\cos^2\theta)$ 
gives the scattering angle $\theta$, which is 
approximately equal to the inclination angle of the binary $\theta_{\rm v}$
as the jet is aligned with the rotational axis of the binary (Figure \ref{figure:x}).
Since the intensity also depends on the angle,
we expect an anticorrelation between the X-ray intensity and polarization degree.
The estimate of inclination angle from polarization degree gives us a test of our model since it is also measurable from the ratio of GW polarizations $h_{+}/h_{\times}=(1+\cos^2\theta)/2\cos\theta$ with an accuracy of $\sim 5\ (2)$ degrees for a NS-NS (BH-NS) system \citep{Arun:2014ysa}.
Second the X-ray polarization angle on the sky determines another jet direction besides $\theta$.
This angle is degenerate with binary orbital phase in the gravitational wave signal
without higher harmonics \citep{O'Shaughnessy:2013vma}.
Thus the X-ray polarization can improve the measurement of parameters\footnote{Note that none of the X-ray detectors and missions in the paper has polarization sensitivity. Hence this particular aspect would have to be addressed independently by other future detectors and missions.}.

Recently, \citet{Yang+15} reported the discovery of near-infrared bump with luminosity $L\sim10^{41}$ erg s$^{-1}$ that is significantly above the regular decaying afterglow in GRB 060614. 
The plateau emission with $L_{\rm iso, pl}=4\times10^{44}$ and $t_{\rm inj}\sim10^5$ s was detected in this event \citep{KI15}.
Using these values and other parameters, $\eta=0.1$, $\theta_{\rm j}=4^\circ$ and $t=12$ days, the estimated luminosity of the plateau activity-powered macronova, $L\sim10^{41}$ erg s$^{-1}$, is consistent with the observed one. 

\acknowledgments
We are grateful to the anonymous referee for helpful comments. 
We would also like to thank Yi-Zhong Fan, Kenta Hotokezaka, Kazumi Kashiyama, Hiroyuki Nakano, Ehud Nakar, Tsvi Piran and Takanori Sakamoto for fruitful discussions. 
This work is supported by KAKENHI 24103006 (S.K., K.I., TN), 24000004, 26247042, 26287051 (K.I.), 15H02087 (TN).

\end{document}